\documentclass[11pt]{article}
\usepackage{times}
\usepackage{cospar}
\usepackage[sectionbib]{natbib}
\usepackage{url}

\usepackage{graphicx}
\usepackage[figuresright]{rotating}

\title{{\it ASCA} AND {\it RXTE} OBSERVATIONS OF NON-THERMAL X-RAY 
EMISSION FROM GALACTIC SUPERNOVA REMNANTS: G156.2$+$5.7}

\author{T.G. Pannuti$^{1,2}$ and G.E. Allen$^{1}$}

\begin{document}

\maketitle

\begin{center}
{\it $^1$MIT Center for Space Research, 77 Massachusetts Avenue,
Cambridge, MA 02139, USA}\\
{\it $^2$California Institute of Technology, SIRTF Science Center,
MS220-6, Pasadena, CA 91125, USA}
\end{center}

\begin{abstract}
We are conducting a survey of Galactic shell-type supernova remnants 
(SNRs) known or suspected to possess non-thermal components to their X-ray 
emission using new and archived observations made with such X-ray 
satellites as $\it{ROSAT}$, $\it{ASCA}$, $\it{RXTE}$, $\it{Chandra}$ and 
${\it XMM-Newton}$. This research is intended to probe the 
phenomenon of cosmic-ray acceleration by Galactic SNRs and estimate
the maximum energy of cosmic-ray electrons accelerated by these sources.
To illustrate this work, we examine the X-ray spectrum of the northwestern
rim of an SNR suspected to have a non-thermal component to its X-ray
emission, G156.2$+$5.7 (RX J04591$+$5147), over the energy range of 
$\approx$0.7-12.0 keV using observations made by the {\it ASCA}~GIS and 
the {\it RXTE}~PCA. We compare fits made to the non-thermal component 
using two models, a simple power law and $\it{SRCUT}$. Both models give 
acceptable fits: the photon index derived from the fit made with the 
power law model ($\Gamma$ = 2.0$^{+0.2}_{-0.5}$) is comparable to values 
obtained for the bright rims of other SNRs with hard X-ray spectra.
Using the $\it{SRCUT}$ model, we derive a value of 2.42$^{+0.24}_{-0.23}$
$\times$ 10$^{17}$ Hz for the cutoff frequency $\nu$$_\mathrm{cutoff}$: 
based on this value and assuming a mean magnetic field strength of 14 
$\mu$G, we estimate the cutoff energy $E$$_\mathrm{cutoff}$ of cosmic-ray 
electrons accelerated by G156.2$+$5.7 to be $\approx$ 32 TeV. This
energy value is well short of the ``knee" feature of the cosmic-ray spectrum.  
\end{abstract}

\section*{INTRODUCTION}

In recent years the study of non-thermal X-ray emission from Galactic
supernova remnants (SNRs) has 
attracted a considerable amount of interest within the astronomical 
community. Extensive research in this field commenced with the discovery 
that the X-ray emission from the Galactic SNR SN 1006 is predominately
non-thermal (Koyama et al. 1995, Allen et al. 2001).
Since this discovery, the number of Galactic SNRs which are known 
to possess spectra dominated by non-thermal X-ray emission has steadily 
increased and currently includes G347.3-0.5 (Koyama et al. 1997, Slane et al. 
1999, Pannuti et al. 2003), G266.2-1.2 (Slane et al. 2001) and G28.6-0.1 
(Bamba et al. 2001, Koyama et al. 2001) as well as SN 1006. Generally 
speaking, all of these Galactic SNRs are X-ray luminous, radio weak 
sources with shell-type morphologies expanding into low density ambient 
media. It is generally believed that the 
non-thermal X-ray emission is in fact synchrotron emission from 
electrons accelerated along the shock front of the SNR (e.g., Reynolds 
1996, 1998; Reynolds and Keohane 1999), although others have argued that 
for at least some SNRs, the emission is instead produced by non-thermal
bremsstrahlung (e.g. Vink and Laming 2002).
\par
If the observed non-thermal emission does in fact have a synchrotron
origin, it is possible to estimate the maximum energy of 
cosmic-ray electrons accelerated by these SNRs through an analysis of
this emission. While it is currently believed that Galactic SNRs
accelerate cosmic-ray particles to the well-known ``knee" feature of the
cosmic-ray spectrum ($E$$_\mathrm{knee}$ $\approx$ 3000 TeV), only a scant
amount of data is available to test this hypothesis. To probe this issue
in more detail, we are conducting 
an X-ray survey of a sample of Galactic SNRs using new and archived
observations made with such satellites as the {\it R\"{o}ntgensatellit} 
({\it ROSAT}), the Advanced Satellite for Cosmology and Astrophysics ({\it 
ASCA}), the Rossi X-ray Timing Explorer ({\it RXTE}), the X-ray Multiple 
Mirror Mission ({\it XMM-Newton}) and the {\it Chandra} X-ray Observatory. 
We are extracting emission spectra from the X-ray bright rims of Galactic 
SNRs known or suspected to possess non-thermal components to their X-ray 
emission. To illustrate this work, in this paper we consider one of the 
sources in our sample, G156.2$+$5.7 (RX J04591$+$5147): this SNR is 
suspected to 
possess both thermal and non-thermal components to its X-ray emission  
(Yamauchi et al. 1993, Yamauchi et al. 1999, Tomida 1999). We analyze 
observations made of the northwestern rim of this SNR using the Gas Imaging 
Spectrometer (GIS) aboard {\it ASCA} and the Proportional Counter Array 
(PCA) aboard {\it RXTE}. 
A detailed summary of these observations is provided in Table 1. In 
Figure 1, we present an X-ray image of G156.2$+$5.7 that was prepared using 
six pointed observations made with the 
Position Sensitive Proportional Counter (PSPC) aboard $\it{ROSAT}$.  

\begin{center}
\begin{table}[t]
\vspace{-0.3cm}
\caption{{\it ASCA}~and {\it RXTE}~Observations of the Northwestern
Rim of G156.2$+$5.7 (RX J04591$+$5147)}
\begin{tabular}{lccccccc}
\hline
\\
& & & & & & Effective & Range of \\
& & & R.A. & Decl. & Radius of & Exposure & Sampled \\
Satellite and & SeqID or & Observation & (h m s) & ($^{\circ}$ m s) & 
Field of View & Time & Energies \\
Instrument & ObsID & Date & (J2000.0) & (J2000.0) & (arcmin) & (Seconds)
& (keV) \\
\hline
{\it ASCA}~GIS2+GIS3 & 50011000 & 1993 Sep 25 & 04 59 14 & $+$52 
29 56 & 22 & 32964 & 0.7--2.5\\
{\it ASCA}~GIS2+GIS3 & 57059000 & 1999 Mar 5 & 04 55 36 & $+$52
06 52 & 22 & 42570 & 0.7--2.5\\
{\it RXTE}~PCA & 50150-01 & 2002 Jan 25 & 04 56 45 & $+$52 21 00 & 60 & 
16160 & 3.0-12.0\\
\hline
\end{tabular}
\end{table}  
\end{center}

\section*{THE GALACTIC SNR G156.2$+$5.7 (RX J04591$+$5147)}
The Galactic X-ray SNR G156.2$+$5.7 (RX J04591$+$5147) was discovered 
during the $\it{ROSAT}$ All-Sky Survey and first described by Pfeffermann 
et al. (1991). Using a Sedov model to fit its X-ray spectrum, those authors 
derived an age of 26000 yr for G156.2$+$5.7, a distance of 3 kpc (which
is supported by an independent distance estimate of 1-3 kpc based on the 
column density toward this source as well as HI observations) and an 
extremely low ambient density of 0.01   
cm$^{-3}$. Assuming a distance of 3 kpc, the X-ray flux from this 
SNR given by Pfeffermann et al. (1991) (1.9$\times$ 10$^{-10}$ ergs 
cm$^{-2}$ sec$^{-1}$ between 0.1 and 2.4 keV) indicates a luminosity of
2.0 $\times$ 10$^{35}$ ergs sec$^{-1}$, placing this source among the 
ten most luminous known Galactic X-ray SNRs. Complementary radio 
observations of G156.2$+$5.7 presented by Reich et al. (1992) indicated a 
modest flux density of 4.2$\pm$0.1 Jy at 1 GHz and an extremely low 
surface brightness of 5.8 $\times$ 10$^{-20}$ ergs cm$^{-2}$ sr$^{-1}$.
Reich et al. (1992) also noted that the radio emission from the
rims of this SNR is highly polarized, consistent with a synchrotron
origin. These characteristics -- high X-ray luminosity but low radio luminosity,
low ambient medium density, and polarized radio emission from the rims
of the SNR -- are generally in common with the other Galactic SNRs known 
to possess X-ray spectra dominated by non-thermal emission. 
Additional X-ray observations of G156.2$+$5.7 have
been presented using observations made with $\it{ROSAT}$,
$\it{Ginga}$~(Yamauchi et al. 1993) and $\it{ASCA}$~(Yamauchi et al.
1999). Both of those papers commented on the harder component to the X-ray
spectrum of this SNR, with Yamauchi et al. (1999) deriving a photon index 
of $\Gamma$=2.1$^{+0.3}_{-0.2}$ for a power-law fit to this component over 
the $\it{ASCA}$ energy band. This index is comparable to the indicies 
derived by power-law fits to the hard components of the $\it{ASCA}$~X-ray 
spectra from other Galactic SNRs.

\section*{OBSERVATIONS OF G156.2$+$5.7}
\subsection*{ASCA GIS2+GIS3 Observations}
Both the GIS and the Solid-State Imaging Spectrometer (SIS) aboard 
$\it{ASCA}$ (Tanaka et al. 1994) observed two different 
positions along the northwestern rim of
G156.2$+$5.7, but becuase the GIS has a larger field of view than the
SIS we only consider data from that instrument in this work. 
The GIS consists of two units (denoted as GIS2 and
GIS3) of imaging gas scintillation proportional counters with a sealed-off
gas cell equipped with an imaging phototube. The GIS 
is sensitive over the energy range of approximately 0.7 -- 10 keV and has 
an energy resolution of $\Delta$$E$/$E$ = 0.078 at 6 keV (Ohashi et al. 
1996, Makishima et al. 1996). The reduction process filtered data for 
elevation 
\begin{center}
\begin{minipage}{82mm}
\includegraphics[width=82mm]{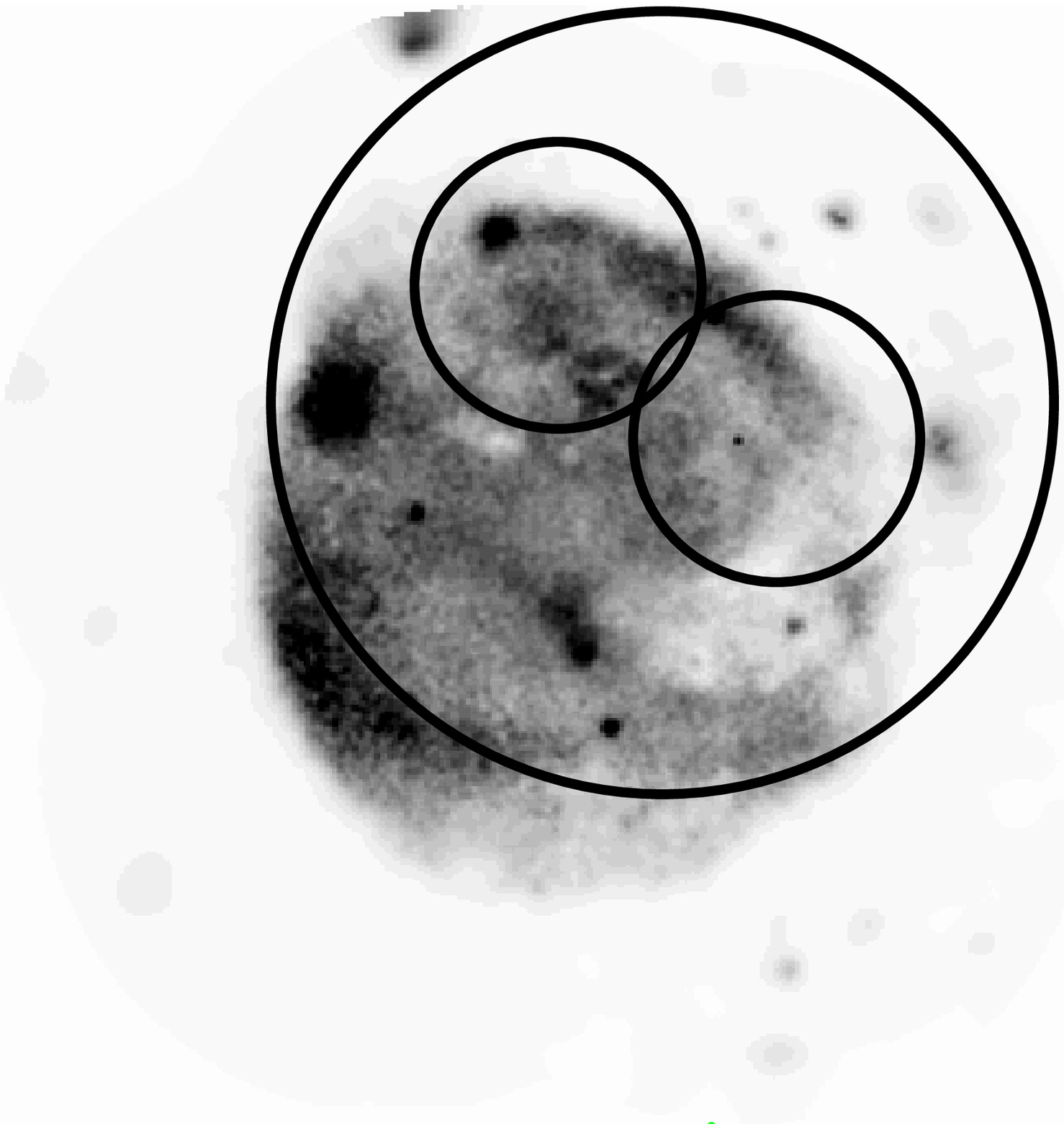}
{\sf Fig. 1. {\it ROSAT} PSPC X-ray image of G156.2$+$5.7 depicting 
emission over the energy range of 0.4-2.0 keV (courtesy of Steven Snowden). 
The two small circles represent the fields of view of the $\it{ASCA}$ GIS 
observations, while the large circle represents the field of view of the 
$\it{RXTE}$ PCA observation (see Table 1 and ``Observations of 
G156.2+5.7.")}
\end{minipage}
\hfil\hspace{\fill}
\begin{minipage}{80mm}
\includegraphics[width=80mm]{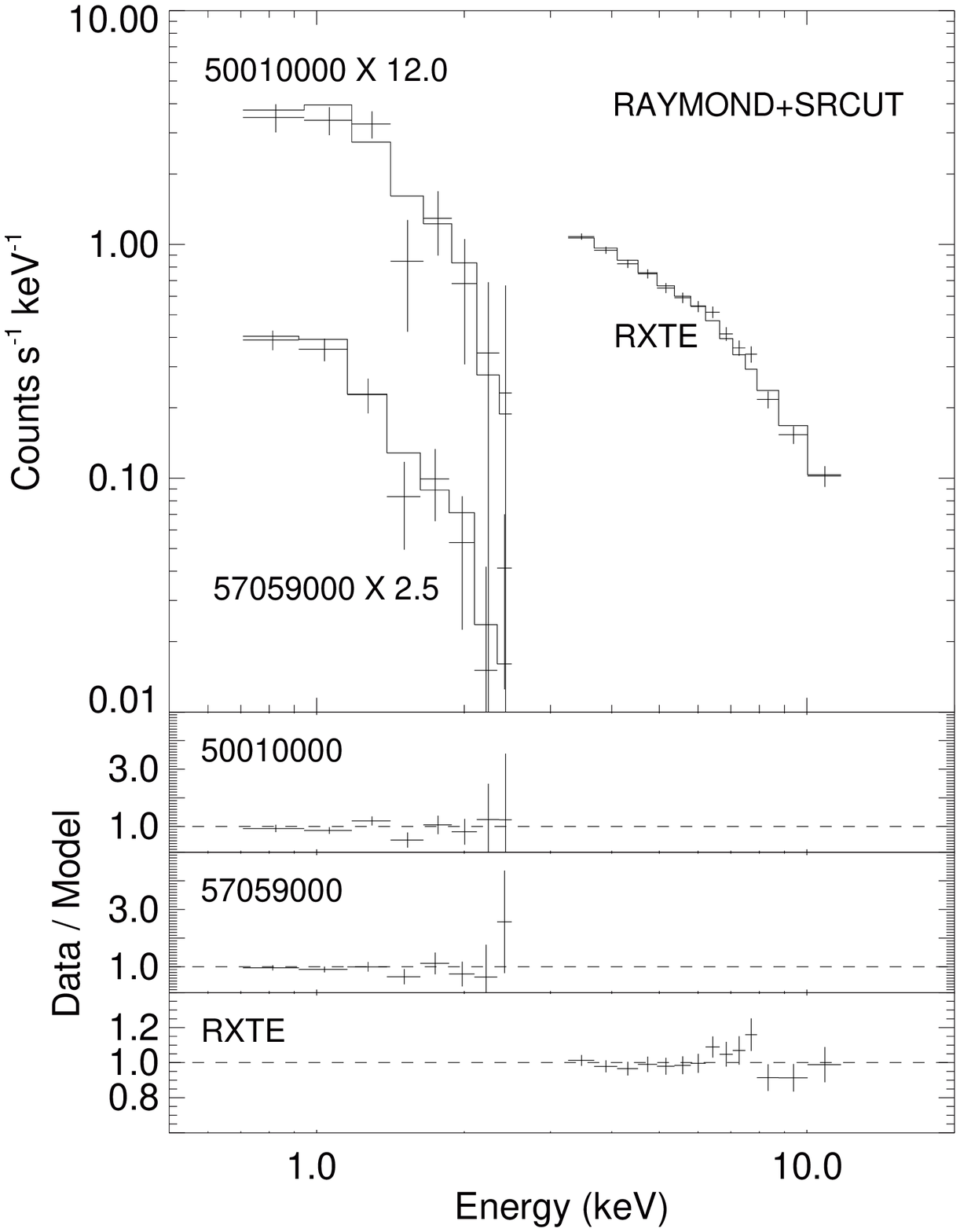}
{\sf Fig. 2. X-ray spectra of the northwestern rim of G156.2$+$5.7 as
observed by the {\it ASCA}~GIS and the {\it RXTE}~PCA (see Table 1 and
``Results").}
\end{minipage}
\end{center}
angle, stable pointing directions, the South Atlantic Anomaly
and cut-off rigidity. The data were also screened for characteristics of
the GIS internal background and screening based on event locations and
rise time. The standard script ``ASCASCREEN" was run to accomplish
this reduction. To boost the signal-to-noise, for both pointings we
combined the spectra observed by the GIS2 and GIS3 units  
and considered these combined spectra in our analysis.

\subsection*{RXTE PCA Observations}

Because of the rather large angular extent of G156.2$+$5.7 ($\approx$
110 arcminutes, Green 2001), it was not possible to observe the entire
SNR with a single pointing of the PCA, so we instead observed just the
luminous northwestern rim. The PCA is a spectrophotometer comprised of 
an array of five coaligned proportional counter units that is sensitive
to photons that have energies between approximately 2 and 60 
keV. The energy resolution $\Delta$$E$/$E$ of the array is 0.18 at 6 keV, 
and the maximum on-axis collecting area is about 6000 cm$^2$ at 9.72 keV.
The PCA data were screened to remove the time intervals
during which (1) one or more of the five proportional counter units
is off, (2) G156.2$+$5.7 is less than 10$^{\circ}$ above the limb of
the Earth, (3) the background model is not well-defined and (4) the
pointing direction of the detectors is more that 0$^{\circ}$.02 from   
the nominal pointing direction in either right ascension or declination.

\section*{RESULTS}

We first considered the two $\it{ASCA}$ GIS observations of the
northwestern rim: these observations were sensitive to thermal 
emission from G156.2$+$5.7 and we used the data from these 
observations to model this type of emission. We point out that the point 
source AX J0500$+$5238 lies along the northwestern rim and 
is believed to be a background galaxy unrelated to G156.2$+$5.7
(Tomida 1999, Yamauchi et al. 1999). We extracted spectra
for two regions along this rim as sampled by the two $\it{ASCA}$ GIS 
pointings (excluding flux from AX J0500$+$5238) and fit
the spectra using a thermal $\it{RAYMOND}$ model (Raymond and Smith
1977) with abundances frozen to solar values. Photoelectric absorption 
along the line of sight was fit using the $\it{WABS}$ model, which is 
based on the Wisconsin cross-sections (Morrison and McCammon 1983) and 
relative elemental abundances as described by Anders and Ebihara (1982).
For the spectra extracted from both regions we obtained acceptable fits 
using $\it{WABS}$$\times$$\it{RAYMOND}$~in both cases. In the case of
the SeqID 50011000 observation, our fit values were $\it{kT}$ = 
0.48$^{+0.19}_{-0.20}$ keV and $N$$_\mathrm{H}$ = 0.47$^{+0.25}_{-0.19}$ 
$\times$ 10$^{22}$ cm$^{-2}$ with $\chi$$^2$/degrees of freedom = 42.6/149, 
while for the SeqID 57059000 observation our fit values were $\it{kT}$ = 
0.57$^{+0.13}_{-0.17}$ keV and $N$$_\mathrm{H}$ = 0.24$^{+0.15}_{-0.18}$ 
$\times$ 10$^{22}$ cm$^{-2}$ 
with $\chi$$^2$/degrees of freedom = 71.6/149. We therefore conclude
that the thermal emission from the northwestern rim of G156.2$+$5.7. may 
be adequately modeled using the a column density and a temperature 
corresponding to the average values for these two parameters derived 
from these two fits; that is, assuming $N$$_\mathrm{H}$ = 0.36 $\times$
10$^{22}$ cm$^{-2}$ and $\it{kT}$ = 0.52 keV. 
\par
We next considered the $\it{RXTE}$~PCA observation of the northwestern
rim of G156.2$+$5.7: because of the large field of view of the PCA, this
observation sampled flux from G156.2$+$5.7 as well as AX J0500$+$5238 
and a second background galaxy, 3C 130. To properly account for emission
from these two galaxies, we first extracted and modeled a GIS spectrum 
for AX J0500$+$5238 and fit the spectrum over the energy range of 0.7--6.0 
keV using a power law and the $\it{WABS}$ model for photoelectric absorption. 
We obtained an acceptable fit ($\chi$$^2$/degrees of freedom = 90.5/448)
to the spectrum with a photon index $\Gamma$ = 2.0$^{+1.1}_{-0.7}$
and a column density $N$$_\mathrm{H}$ = 0.57$^{+0.94}_{-0.57}$ $\times$ 
10$^{-22}$ cm$^{-2}$ (both broadly consistent with Tomida 1999 and 
Yamauchi et al. 1999). In the case of 3C 130 (which was not sampled by the 
$\it{ASCA}$~GIS observations), Hardcastle (1998) obtained an acceptable 
fit to the spectrum of this source using the $\it{RAYMOND}$ model 
($\it{kT}$ = 2.9$^{+9}_{-2}$ keV) and a column density $N$$_\mathrm{H}$ = 
0.9$^{+0.5}_{-0.2}$ $\times$ 10$^{22}$ cm$^{-2}$, assuming a redshift 
of 0.109. We adopt this model in our fitting procedure to describe the 
X-ray emission from 3C 130. 
\par
To model the non-thermal component to the observed X-ray emission 
from the northwestern rim of G156.2$+$5.7, we considered two 
different models: a simple power law model and the $\it{SRCUT}$ model
(Reynolds 1998, Reynolds and Keohane 1999). This latter model describes
a synchrotron spectrum from a power-law distribution (with an exponential
cutoff) of electrons in a uniform magnetic field. The photon
spectrum is itself a power-law, rolling off more slowly than exponential 
in photon energies. 
The $\it{SRCUT}$ model can be used to estimate the maximum energy of 
electrons accelerated by the SNR: we assume that the relativistic 
electron energy spectrum $N$$_\mathrm{e}$($E$) may be expressed as 
$N$$_\mathrm{e}$ ($E$) = $K$ $E$$^{-\Gamma}$ $e$$^{-E/E_\mathrm{cutoff}}$, 
where $K$ is a normalization constant derived from the observed
flux density of the region of the SNR at 1 GHz, $\Gamma$ is defined as
2$\alpha$+1 (where $\alpha$ is the radio spectral index) and
$E$$_\mathrm{cutoff}$ is the maximum energy of the accelerated cosmic-ray
electrons. A crucial advantage of this model is that a resulting fit may 
be compared with two observable properties of an SNR, namely its flux 
density at 1 GHz and $\alpha$. Also, one of the fit parameters for this model 
is the cutoff frequency $\nu_\mathrm{cutoff}$ of the electron synchrotron 
spectrum, defined as the frequency at which the flux has dropped by a 
factor of 10 from a straight power law. This frequency may be expressed as
\begin{equation}
\nu_\mathrm{cutoff} \approx 1.66 \times 10^{16} 
\left( \frac{B_\mathrm{\mu G}}{10~\mu G} \right) \left( 
\frac{E_\mathrm{cutoff}}{10~\mbox{TeV}} \right)^2
\quad \mbox{Hz,} \quad
\end{equation}
where $B_\mathrm{\mu G}$ is the magnetic field strength of the SNR in
$\mu$G, assuming the electrons are moving perpendicular to the magnetic 
field. By using the value for $\nu_\mathrm{cutoff}$ returned by  
$\it{SRCUT}$, an estimate for the maximum energy $E$$_\mathrm{cutoff}$ for 
the shock-accelerated electrons may be calculated. 
\par
We therefore modeled the $\it{RXTE}$~PCA spectrum using a four component
model: a component for the thermal emission from the rim, a component
for the emission from AX J0500$+$5238, a component for the emission from
3C 130 and finally a component for the nonthermal emission from the rim.
For the first three components, we adopt the previously-described models 
and parameter values, while for the final component we separately used and
compared a power law model and an $\it{SRCUT}$ model. Both of these
models give acceptable fits to the non-thermal component of the spectrum:
in the case of the power law model, for the photon index we obtain
$\Gamma$ = 2.0$^{+0.2}_{-0.5}$ ($\chi$$^2$/degrees of freedom =
16.2/15 = 1.08). Our value for the photon index is similar to 
values found for the bright rims of other SNRs that feature 
non-thermal spectra, such as G266.2$-$1.2 and G347.3$-$0.5 (Slane 
et al. 1999; 2001). This value is also comparable to the value obtained by 
Yamauchi et al. 1999 ($\Gamma$ = 2.1) in their analysis of $\it{Ginga}$
observations of G156.2$+$5.7. Using the $\it{SRCUT}$ model also yields
an acceptable fit ($\chi$$^2$/degrees of freedom = 19.9/14 = 1.42) with
values of 0.66$\pm$0.05 and 1.58$^{+0.15}_{-0.14}$ Jy for 
$\alpha$ and the normalization, respectively: both 
of these values are broadly consistent with the results of Reich et al.
(1992). It is very remarkable that using the $\it{SRCUT}$ model yields a
fit which closely approximates the known radio properties of this SNR 
(note that the value for the normalization is almost half the
total flux density at 1 GHz of G156.2$+$5.7 and reflects the bilateral
radio morphology of this SNR).
For the cutoff frequency parameter of this fit, we obtain
$\nu$$_\mathrm{cutoff}$ = 2.41$^{+0.22}_{-0.23}$ $\times$ 10$^{17}$ Hz:
assuming that the mean magnetic field strength of G156.2$+$5.7 is 14
$\mu$G (Reich et al. 1992) and using Equation (1), we estimate that the 
maximum energy of cosmic-ray electrons accelerated along the northwestern 
rim of G156.2$+$5.7 is $\approx$ 32 TeV, well short of the ``knee" 
feature of the cosmic-ray spectrum. In Figure 2, we present fits to the 
two $\it{ASCA}$ GIS2+GIS3 spectra and the $\it{RXTE}$~PCA spectrum that
we have extracted for this analysis, using the $\it{SRCUT}$ model to 
fit the non-thermal X-ray emission seen in the $\it{RXTE}$~PCA spectrum.
Additional X-ray observations of the northwestern rim of G156.2$+$5.7 with 
improved sensitivity are required to more rigorously constrain its
spectral properties. 

\section*{COMPARISONS WITH OTHER GALACTIC SNRS}

In Table 2, we present a comparison of the gross properties of 
five Galactic SNRs (SN 1006, G28.6$-$0.1, G266.2$-$1.2 and G347.3$-$0.5,
as well as G156.2$+$5.7) with X-ray spectra which feature  
non-thermal emission. All of the SNRs in this class are X-ray luminous but 
radio faint and are expanding into regions of low ambient density ($n$ 
$\leq$ 0.2 cm$^{-3}$): it appears that searches for other SNRs in this 
class should concentrate on sources expanding into a low density media. 
However, we point out that two other SNRs which are known to possess 
non-thermal components to their X-ray spectra, Cas A and RCW 86, are
expanding into high density ambient media. Therefore, the density of 
the ambient media is not the only factor in determining the strength of 
the non-thermal component to X-ray emission from SNRs. No clear evidence
exists that any of these five SNRs are currently accelerating cosmic-ray
electrons to the ``knee" of the cosmic-ray spectrum: in fact, estimates
for $E$$_\mathrm{cutoff}$ for each of these SNRs often fall short by an 
order of magnitude or more (Reynolds and Keohane 1999). Additional 
observations of Galactic SNRs using the present generation of X-ray 
observatories are required in order to analyze the phenomena of 
non-thermal X-ray emission
and cosmic-ray acceleration by SNRs. 

\begin{table}[t]
\vspace{-0.3cm}
\caption{Gross Properties of Galactic SNRs with Non-Thermal-Dominated
X-ray Spectra}
\begin{tabular}{lccccc}
\hline
\\
SNR & SN 1006 & G28.6$-$0.1 & G156.2$+$5.7 & G266.2$-$1.2 & G347.3$-$0.5\\
\hline
Other Names & -- & AX J1843.8$-$0352 & RX J04591$+$5147 & 
RX J0852.0$-$4622 & RX J1713.7$-$3946 \\
Distance (kpc) & 2.2$^1$ & 7$^2$ & 1.3$^3$ & 1-2$^4$ & 6$^5$ \\
Age (yr) & 997$^1$ & $\leq$2700$^2$ & 15000$^3$ & 
$\approx$14000$^{\dagger}$ & $\approx$8000$^6$\\
Size (arcmin) & 30$^7$ & 13$\times$9$^7$ & 110$^7$ & 120$^7$ & 
65$\times$55$^7$\\
$S$$_\mathrm{1.4~GHz}$ (Jy) & 19$^7$ & 3(?)$^7$ & 4.2$\pm$0.1$^8$ & 
50(?)$^7$ & 4$\pm$1$^6$\\
$\alpha$ ($S$$_{\nu}$ $\propto$ $\nu$$^{-\alpha}$) & 0.6$^7$ & -- & 
0.5$^7$ & 0.3(?)$^7$ & $\approx$0.55$^9$\\
$\Gamma$ & 2.1$^{10}$ & 2.1$^2$ & 2.1$^3$ & 2.6$^4$ & 2.4$^5$ \\
$B$ ($\mu$G) & 9$^{11}$ & -- & 12-16$^{8}$ & -- & 
150($^{+250}_{-80}$)~$^9$ \\
$n$ (cm$^{-3}$) & $\approx$0.1$^{10}$ & $\leq$0.2$^2$ & $\approx$0.2$^3$
& 0.05$^{\dagger}$ & $\approx$0.05-0.07$^9$ \\
Progenitor Type & Ia$^1$ & -- & -- & II$^4$ & II$^5$ \\
\hline
\end{tabular}
References: $^1$ -- Winkler et al. 2002, $^2$ -- Bamba et al. 2001, 
$^3$ -- Yamauchi et al. 1999, $^4$ -- Slane et al. 2001, $^5$ -- Slane
et al. 1999, $^6$ -- Ellison et al. 2001, $^7$ -- Green 2001, $^8$ -- 
Reich et al. 1992, $^9$ -- Pannuti et al. 2003, $^{10}$ -- Allen et al.
2001, $^{11}$ -- Dyer et al. 2001. $^{\dagger}$ -- Calculated by Pannuti 
et al. 2003 using expressions derived by Slane et al. 2001 and assuming a 
distance of 1.5 kpc, a filling factor for a sphere of 1/4 and 
an explosion kinetic energy of 10$^{51}$ ergs. 
\end{table}

\section*{ACKNOWLEDGEMENTS}

We thank the referee for useful comments which helped enhance
the quality of this paper. We are very grateful to Steven Snowden for 
preparing 
the {\it ROSAT}~PSPC image of G156.2$+$5.7 for the purposes of this work.
T. G. P. acknowledges a useful discussion with W. Reich regarding the radio 
properties of G156.2$+$5.7. T. G. P. also acknowledges support for 
this work from NASA LTSA grant NAG5-9237.

\vspace{0.2cm}
Email address of T.G. Pannuti: tpannuti@space.mit.edu

\end{document}